# A Metallurgical Inspection Method to Assess the Damage in Performance-Limiting Nb$_3$Sn Accelerator Magnet Coils

Alice Moros, Mickael Denis Crouvizier, Ignacio Aviles Santillana, Stefano Sgobba, Susana Izquierdo Bermudez, Nicholas Lusa, Jose Ferradas Troitino, Attilio Milanese, Ezio Todesco, Arnaud Devred, Giorgio Ambrosio, Maria Baldini, Paolo Ferracin, Jesse Schmalzle, and Giorgio Vallone

*Abstract*—The design and production of Nb$_3$Sn-based dipole and quadrupole magnets is critical for the realization of the High-Luminosity Large Hadron Collider (HL-LHC) at the European Organization for Nuclear Research (CERN).

Nb$_3$Sn superconducting coils are aimed at enhancing the bending and focusing strengths of accelerator magnets for HL-LHC and beyond.

Due to the brittle nature of Nb$_3$Sn, the coil fabrication steps are very challenging and require very careful QA/QC. Flaws in the Nb$_3$Sn filaments may lead to quenches, and eventually, performance limitation below nominal during magnet testing.

A novel inspection method, including advanced non-destructive and destructive techniques, was developed to explore the root-causes of quenches occurring in performance-limiting coils.

The most relevant results obtained for MQXF coils through this innovative inspection method are presented. This approach allows for precise assessment of the physical events associated to the quenches experienced by magnet coils, mainly occurring under the form of damaged strands with transversely broken sub-elements. Coil-slice preparation, micro-optical observations of transverse and longitudinal cross-sections, and a deep etching technique of copper will be illustrated in the present work, with a focus on the results achieved for a CERN coil from a non-conforming quadrupole magnet prototype, and two coils fabricated in the US, in the framework of the Accelerator Upgrade Project (AUP) collaboration, from two different non-conforming quadrupole magnets, respectively.

The results obtained through the proposed inspection method will be illustrated.

*Index Terms*—Metallurgical inspections, HL-LHC, accelerator magnets, MQXF coils, Nb$_3$Sn, quench damage assessment

## I. Introduction

THE development of high-performance Nb$_3$Sn-based magnets capable of generating high field or high-field gradient with conductor peak field in the 11–13 T range represents a critical step for the realization of the high luminosity LHC upgrade at CERN [1],[2].

Nb$_3$Sn is currently the only low-temperature superconductor offering high critical currents at high fields that is available at an industrial scale [3]. For this reason, it is also the best candidate for the magnets of future particle accelerators such as the CERN Future Circular Collider (FCC) [4] and is widely applied for the cable-in-conduit conductors (CICC) of the ITER magnet system [5],[6].

Despite its excellent electrical properties, Nb$_3$Sn is a brittle material and its superconducting properties are sensitive to strain [7]. When subjected to moderately severe loading conditions, it can eventually trigger the breakage of some superconducting filaments (or aggregated filaments within the strands), thereby hindering current transport and subsequently degrading the magnet performance. The identification of performance-limiting coils during the magnet testing phase led to the need of a step-by-step approach for identifying all possible failures in manufacturing, assembly or operation, in order to further improve manufacturing processes and explore the root causes behind the occurrence of performance limitation below nominal current [8].

The proposed inspection method aims at unveiling the metallurgical causes of the quenches previously localised through voltage taps and quench antennas [9]. It started to be developed for the performance limitation analysis of HL-LHC 11 T dipole magnet coils [9],[10],[11], based on the knowledge inherited from studies on the ITER magnet system [5], [12], [9].

The results obtained on 11 T magnet coils [10] paved the way for the analysis of MQXF performance-limiting coils. In this context, the inspection method was consolidated and optimized. It is of paramount importance to check the possible presence of cracks in the Nb$_3$Sn filaments. Cracks can be classified into three categories: 1) radial, entirely or partially crossing the Nb$_3$Sn region from the filament core to the periphery, 2) circumferential and 3) transverse with respect to the longitudinal axis of the superconducting filaments or sub-elements. They can lead to performance degradation if they propagate during magnet operation. Significant work has been done to establish a relationship between transverse compressive stresses

Manuscript receipt and acceptance dates will be inserted here.

Alice Moros, Mickael Denis Crouvizier, Ignacio Aviles Santillana, Stefano Sgobba, Susana Izquierdo Bermudez, Nicholas Lusa, Jose Ferradas Troitino, Attilio Milanese, Ezio Todesco, and Arnaud Devred are with the European Organization for Nuclear Research (CERN), 1211 Geneva 23, Switzerland (e-mails: alice.moros@cern.ch, mickael.denis.crouvizier@cern.ch).

Giorgio Ambrosio and Maria Baldini are with the Fermi National Accelerator Laboratory, Batavia, IL 60510 USA.

Paolo Ferracin and Giorgio Vallone are with Lawrence Berkeley National Laboratory, Berkeley, CA 94720-8203 USA.

Jesse Schmalzle is with Brookhaven National Laboratory, Upton, NY 11973 USA.

representative of magnet operational conditions and radial crack occurrence [13],[14],[15],[16]. In a combined Focused Ion Beam (FIB)-Scanning Electron Microscopy (SEM) analysis performed at CERN, such cracks were found to be 20-30 µm deep, mainly parallel to the longitudinal axis of the strands and not expected to create a discontinuity through the current-carrying superconducting phase [9].

Transverse cracks, resulting in severe breakage of $Nb_3Sn$ filaments (or in aggregates of filaments in the strand sub-elements), are the main root-cause to be investigated for critical current ($I_c$) degradation, as they reduce the overall cross-section of the superconducting phase [16],[9].

The deep Cu etching technique applied in this study, based on previous research works [12],[17],[18] and described in detail in [9], was found to be very effective for the clear identification of transverse cracks within coils limiting the performance of quadrupole magnet prototypes from CERN (MQXFBP) and from the US-AUP collaboration (MQXFA). This paper presents an innovative application of the above technique to large-size coil samples. An extensive study of large coil volumes through systematic cuts and metallurgical inspections allowed physical events associated to the quenches experienced by performance-limiting coils to be univocally identified.

## II. SAMPLES AND METHOD

An overview of the inspected coils is given in Table 1. All presented coils were limiting the performance of the three magnets MQXFBP1, MQXFA07 and MQXFA08. The quench associated to the performance degradation was located in a central region of the coil (MQXFBP1) and in the coil head at the Connection Side or CS (MQXFA07 and MQXFA08). CS is used to indicate the coil end with the powering system. On the opposite side of the CS is the Non-Connection Side or NCS. Specific information about the history of the examined coils and general coil manufacturing features and performance can be found in [9],[19],[20],[21],[22],[24],[25], [26].

TABLE I
INSPECTED PERFORMANCE-LIMITING COILS

| Coil id. | Manufacturer | From magnet | Coil length (m) | Strand layout | Coil quench location |
|---|---|---|---|---|---|
| CR 108 | CERN | MQXFBP1 | 7.2 | RRP 108/127 | Central region |
| AUP 214 | US HL-LHC AUP[a] | MQXFA07 | 4.5 | RRP 108/127 | Head at CS |
| AUP 213 | US HL-LHC AUP | MQXFA08 | 4.5 | RRP 108/127 | Head at CS |

[a]Work shared among a consortium of US laboratories within the HL-LHC AUP: FNAL, LBNL, and BNL.

The applied inspection methods include both non-destructive (NDT) and destructive techniques and allow physical events associated to quenches previously detected by voltage taps and quench antennas to be revealed. NDT relies on the use of X-ray Computed Tomography (CT) to assess the outcome of the coil winding process and evaluate the possible presence of bulged or popped in/out cable strands. Coil-volumes up to 200-300 mm in length can be analysed and reconstructed with non-conventional systems, i.e., high-energy LINAC X-ray CT [9],[10]. The combination of the latter and lower energy micro-CT is very effective to identify Volumes Of Interest (VOIs) to be extracted for the eventual assessment of local damages, such as cracked strands with broken filaments [9]. A destructive approach which consists of cutting samples by Diamond Wire Saw (DWS) is then used for detailed microscopic analyses including preparation of planes by grinding/polishing, optical and SEM observations, and deep etching of copper.

### A. Inspection of Transverse and Longitudinal Cross-Sections

The coil VOIs were extracted by DWS using the Diamond WireTec DWS250 device. This technique allows any artifact coming from the cutting process to be minimised. Also, the coil slices are usually cut with some margin (a few mm) with respect to the plane of interest to further reduce the risk of introducing bias into the subsequent inspections. Once the coil-volume containing the plane(s) of interest is extracted, progressive grinding/polishing steps are performed manually until the 1-µm diamond suspension polishing step. This process takes approx. 8 hours. A final vibratory polishing step is then carried out for 24-48 hours placing the plane of interest on a vibratory polisher (Buehler Vibromet 2) with a colloidal silica suspension. The cutting system and fine preparation by grinding/polishing are described in detail in [9].

This preparation method is used for the inspection of both transverse and longitudinal cross-sections (with respect to the coil main axis). The whole process (cutting and grinding/polishing) is longer for transverse cross-sections as the longitudinal specimens are much smaller. The length of the latter is based on the width of the cut coil-slice (typically ≤ 50 mm).

After the fine polishing preparation, optical microscopy examinations of transverse and longitudinal cross-sections are performed using a digital microscope Keyence VHX-7000 and optical microscope AXIO Imager from Zeiss. Optical observations provide information on the distribution of radial cracks over the coil transverse cross-sections and allow the extent of transverse breakage along the coil axis in the longitudinal planes to be evaluated.

Furthermore, the investigation by optical microscopy of finely prepared samples allows to assess the status of the resin/fibre glass matrix characterising the inspected coil. The possible presence of metal-to-metal cracks, shrinkage cavities, and resin decohesion effects is thus determined. A detailed overview is provided in [9]. Although the role of the resin in terms of performance degradation is not yet clear, its damage is certainly unwanted as it could contribute to displacement and subsequent deformation/breakage of the filaments.

## B. Deep Copper Etching Process

Transverse cracks are of great interest as they are associated with the reduction of transport properties. The first step to clearly identify these crack-events is the removal of copper through an etching process. In fact, the Cu matrix acts like a support for the superconducting strands, so hindering the potential presence of transverse cracks in the aggregates of $Nb_3Sn$ filaments [9]. After a light polishing of the coil-slice, the Cu etching process is carried out in several steps using a solution of $HNO_3$ and water (ratio 1:1), which is homogeneously spread all over the transverse cross-section with a pipette. Optical observations of the coil transverse plane between etching steps allow the amount of copper removed to be monitored. The thickness of the removed copper (typically 600 µm) can be checked by performing a 3D reconstruction of the etched volume using a digital microscope Keyence VHX-7000. Since the prepared coil plane is very close to the previously detected quench location, removal of about 500-600 µm of Cu is usually sufficient to observe broken filaments. Further details about the deep Cu etching process are reported in [9].

Through the removal of copper, fractured filaments are clearly visible when transverse cracks are present entirely breaking the sub-element cross-section. Their aspect was examined in detail by SEM imaging using the SIGMA Field Emission Gun-SEM from Zeiss.

## III. RESULTS AND DISCUSSION

The following section reports some of the most relevant results obtained from the metallurgical inspection of the coils presented in the previous section. Through this inspection, the physical damages behind their performance limitations could be understood.

### A. CERN Coil MQXFB CR108 from Magnet MQXFBP1

Voltage taps and quench antennas signals precisely located the position of the quench in the central part of the coil [21], [26]. Following precise metrological measurements by Linac CT, four specimens (approx. 30 mm thick) were cut around the detected quench points at the layer jump (LJ) side of a central zone approx. 400 mm wide. The LJ side is meant as the left coil half of the inspected plane faced towards the CS: it includes the coil inner layer and the outer layer (the latter being wound on top of the former). The coil half being opposite to LJ is called opposite layer jump (OLJ). Both the LJ and OLJ sides contain Cu wedges separating sectors of Rutherford-type cables in straight portions of the coil. Segmented Titanium pole pieces are located in-between the two coil-halves along the coil length (coil heads excluded). The gaps between Ti pole segments are defined as pole-to-pole transitions. At the coil heads, transitions between Cu wedges and end spacers made of stainless steel with ceramic coating are separating sectors of Rutherford-type cables. These transitions at the coil heads are spaces filled with resin.

At first, the transverse cross-sections of the four coil slices were finely polished to reveal the presence of radial cracks and evaluate the status of the resin. The observed radial microcracks did not always cross fully the $Nb_3Sn$ phase in the sub-elements and clustered mainly in cables of the Midplane Block (MB) of the coil. The resin showed decohesion and metal-to-metal cracks for 3 of the 4 examined specimens. As all slices showed similar results, two representative images for the examination of radial cracks and resin status over a transverse cross-section (LJ side) are shown in Figure 1.

The progressive dissolution of the Cu matrix allowed the identification of strands with broken sub-elements for three of the prepared transverse cross-sections. A very specific damage location was observed: the same and only top strand of the pole turn cable facing the titanium pole at the outer radius of the inner layer.

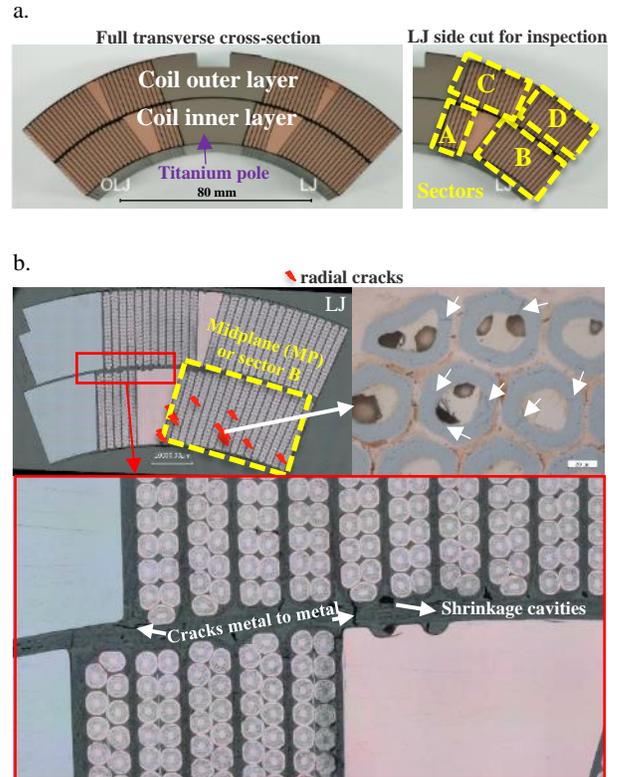

Fig. 1. MQXFBP1 quadrupole magnet coil CR 108: (a) as-prepared coil transverse cross-section displaying the inner/outer layer and the Ti pole (left), and coil LJ side cut for the inspection (right). The yellow dashed frames and the letters A, B, C, D identify the four sectors containing 5, 17, 12, and 16 Rutherford-type cables, respectively. (b) Representative images of the radial cracks observed over the transverse cross-section of coil LJ side cut (top). These cracks are mostly clustered in the cables of the MB or sector (red pins and white arrows). The resin shows shrinkage cavities and metal-to-metal cracks (bottom).

This location is identified in Figure 2 as "pole turn at inner layer" shown along with a SEM close-up of the damaged strand with broken sub-elements. The identified damaged strands are included in a 60 mm long section of the coil (along its main axis) where the limiting quench origin was localised.

SEM analysis of these strands showed the broken sub-elements to have the typical $Nb_3Sn$ intergranular fracture aspect (see Figure 3).

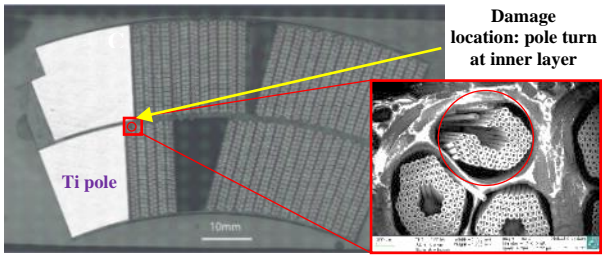

Fig. 2. MQXFBP1 quadrupole magnet coil CR 108: typical aspect of the transverse cross-section after deep Cu etching. The inset in the red frame shows a SEM close-up of the damaged strand with broken filaments at the recurrent position, the pole turn at inner layer (top edge of the cable facing the Ti pole). Portions of intact strands are visible as well.

Longitudinal samples were then extracted for two affected coil slices adjacent to each other. As explained earlier, the inspection of longitudinal sections allows the extent of the fracture along the coil main axis to be assessed.

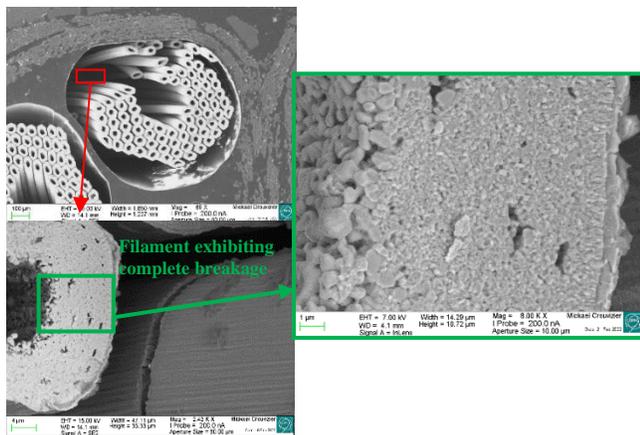

Fig. 3. MQXFBP1 quadrupole magnet coil CR 108: a broken $Nb_3Sn$ filament observed by SEM. The close-up image enclosed by the green frame clearly shows the typical intergranular aspect of $Nb_3Sn$, found in all cases of complete transverse breakage.

For both longitudinal cross-sections, progressive polishing steps were carried out to analyse the extent of the transverse cracks at the midplane of the first and second row of strands within the damaged Rutherford-type cable.

The extent at the level of the first midplane row was found to be approx. 20 mm in one sample and 15 mm in the adjacent one. The major events were at the crests of seven top strands as shown in the representative image of Figure 4.

The close-up at the top right of Figure 4 reports a crack-event resembling a "V" shape fracture (yellow dashed lines): this feature was observed in both specimens and might be related to a flexural stress acting in the coil axial direction following reaction and impregnation [9]. No cracks were detected at the second row midplanes.

Another sample (50 mm thick) was cut in the vicinity of a quench point at the OLJ side of the coil. The same inspection previously described was performed. The Cu etching process revealed damaged strands at the same location as the one observed for the three samples within the LJ side. A longitudinal section was then extracted. Its inspection revealed an extent of strands affected by cracked sub-elements of approx. 30 mm.

The results obtained at this stage suggested that stress concentration on the strand close to the titanium pole at inner layer had likely played a significant role in causing the damage responsible for the quench performance limitation of the coil. The longitudinal extent of crack-events within the central quench zone, considering all the inspected coil slices (4 on the LJ side and 1 on the OLJ side), was about 90 mm.

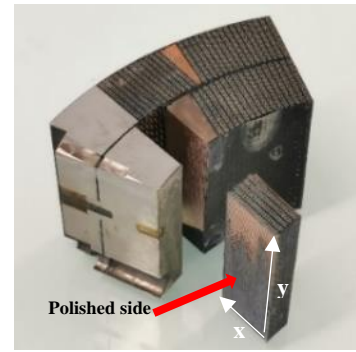

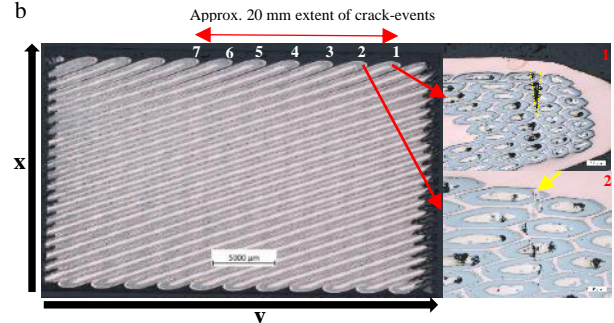

Fig. 4. MQXFBP1 quadrupole magnet coil CR 108: (a) Extracted longitudinal sample where the side to be polished is displayed; (b) Longitudinal cross-section finely polished at the midplane of the first row of strands within the damaged cable (left). The extent of the events is about 20 mm. The major events are located at the crests of seven top strands. Strand numbering starts from the closest strand to the damaged one observed in the transverse cross-section after Cu etching. The two close-up images on the right represent observed crack-events and are highlighted by yellow dashed lines (top), and a yellow arrow (bottom).

Two samples very close to the NCS coil-head were analysed to assess the status of the sub-elements and resin/fibre glass matrix in the region farthest from the damaged central quench zone. In this case, no crack-events or defects in the resin were expected, and the applied fine preparation and deep Cu etching confirmed this hypothesis.

A systematic study with twenty additional samples (50 mm thick) covering one metre coil-length being inspected aimed at verifying the potential presence of broken filaments further away from the central quench region [26]. In three samples, cracked filaments were identified at the same location (top of inner-layer pole turn) as in the quench region. Moreover, the three samples had in common that they were located close to Ti pole-to-pole transitions included in this 1-metre segment. Fig 5 illustrates the positions of the identified damaged strands along the coil: in addition to the three above mentioned (at the coil NCS), another strand with broken filaments was found at the top of inner-layer pole turn in the vicinity of a pole-to-pole transition towards the coil CS.

Longitudinal cross-sections showed the extent of these damages ranging between 15 mm and 20 mm.

Such pole-to-pole transitions are discontinuity points, thus likely creating a risk for stress-concentration. The poles are approx. 400 mm long and there is an axial overlap of approx. 20 mm between the inner layer pole and the outer layer one [26],[9].

The reasons of such preferential occurrence of the damage near the pole-to-pole transitions is being investigated in order to understand at which stage of the coil fabrication or operation they have occurred [26].

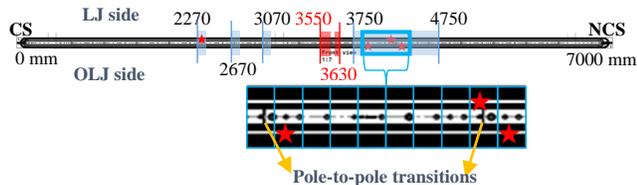
Fig. 5. Scheme of the 7 m long coil CR108 from the magnet MQXFBP1: the light blue rectangles represent the inspected samples, and the red stars display the locations of damaged strands with broken sub-elements. The light blue rectangle on the right shows the analyzed 1-m-segment with its 3 damaged strands (at the top of the inner-layer pole turn). A close-up view of this affected area highlighting the pole-to-pole transitions is shown as well. The red rectangles point out the damages in the areas of quench start localisation.

### B. US Coil MQXFA AUP 214 from Magnet MQXFA07

For the quadrupole magnet MQXFA07, voltage taps and quench antenna data revealed that the origin of the limiting quenches was in the CS ("lead end" in the USA) end zone of coil AUP 214 inner layer [23],[24],[25]. Based on the results of finite element (FE) simulations [24], the possible cause of the limiting quenches was a stress field located between the coil end spacer-Cu wedge (S-W) transition introduced in the previous paragraph. In particular, the two cables facing this transition, at the bottom of the inner layer pole block (PB) and at the top of the midplane block (MB), as illustrated in Fig. 7, had highest probability of strand damage. Priority was thus given to this S-W transition located at the end of the straight section of the coil, near the CS end. The coil sample including the S-W transition is shown in the close-up view of Figure 6 (bottom).

A 10 mm thick coil-slice containing the S-W transition was cut and further split into two parts, separating the LJ and OLJ side to facilitate the analysis (Figure 6). The two coil cross-sections were then embedded in epoxy resin and finely prepared through the method previously described to assess the resin status and identify possible radial micro-cracks. The latter were observed by optical microscopy in the sub-elements all over the LJ and OLJ transverse cross-sections as well as shrinkage cavities through the fibre glass/resin matrix. It was also found that the ceramic coating of the end spacer provides good cohesion with the resin [24]. Copper was then etched away from both cross-sections to identify possible damaged strands. At this stage, no strands seemed to contain any completely broken filaments.

For a better understanding of cable integrity along the whole transition, longitudinal samples were extracted for the analysis of the cables facing the S-W transition on both the LJ and OLJ side (4 samples in total, each including a cable facing the S-W transition). A representative image of the extracted longitudinal samples with the cables of interest (OLJ case) is shown in Figure 7. The yellow dashed lines indicate the cable facing the S-W for both the PB and MB. Also, they highlight the surface prepared for the inspection, i.e., the midplane of the first row of strands belonging to the transition-facing cable.

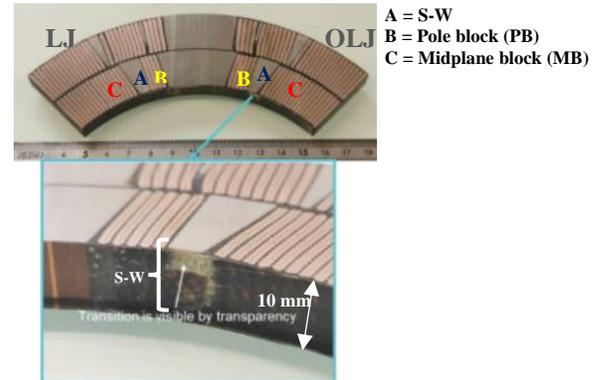
Fig. 6. MQXFA07 quadrupole magnet coil AUP 214: 10 mm thick coil-slice including the S-W transition and cut by DWS (top); close-up image showing that the S-W transition is already visible by transparency through resin (bottom).

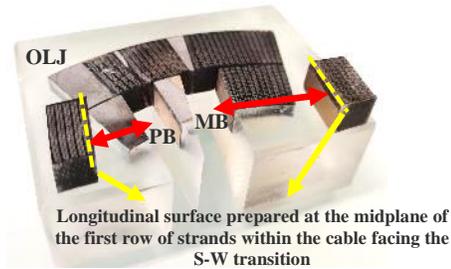
Fig. 7. MQXFA07 quadrupole magnet coil AUP 214: 10 mm thick OLJ side embedded in epoxy resin for the metallurgical inspection. In this case, the extraction of longitudinal samples is illustrated. The red arrows point out the two samples extracted from the PB and MB, respectively. The yellow dashed lines represent the plane to be polished withing the cables facing S-W.

The cracks detected by optical microscopy over the LJ and OLJ longitudinal samples are shown in Figure 8 (top) and Figure 8 (bottom), respectively. Each red mark corresponds to an observed transverse crack going entirely through a $Nb_3Sn$ sub-element for the LJ-MB sample, no cracks were identified. The LJ-PB sample showed instead 532 transverse crack-events confined in a field of approx. 2 mm. Similar results were found for the OLJ specimens: only few crack-events (54) were observed for the OLJ-MB, whereas 728 were counted for the OLJ-PB. For the latter sample with the greatest number of cracks at the midplane of the first row of strands (within the cable facing S-W), also the second row of strands was inspected for possible damage. The same metallurgical inspection did not reveal any crack-event in this case. These observations can be put in parallel with the results of the FE simulations presented in [24],[25]: the location of the dense field of transverse cracks (about 2 mm wide) that is observed at the bottom of the PB for both the LJ and OLJ sides corresponds to the area where a strain peak approached 0.4%. In addition, the inspection of the second row did not reveal any field of cracks: this confirmed that the stress field was localised at the interface between the S-W transition and the 1st row of the Rutherford-type cable facing it.

The exact location of the crack field with respect to the S-W transition appeared to be in correspondence of the interface between the resin and the Cu wedge. Other related details are given in [24],[25].

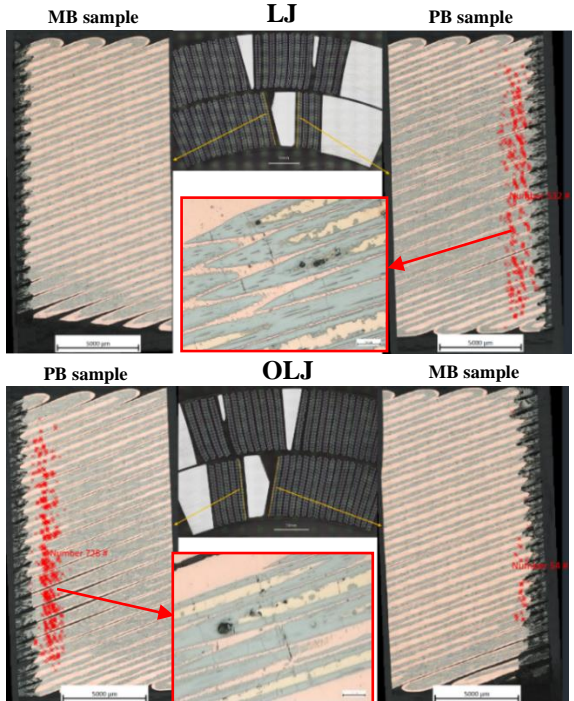

Fig. 8. MQXFA07 quadrupole magnet coil AUP 214: transverse crack-events (red marks) over the longitudinal cross-sections prepared for the LJ-PB and LJ-MB samples (top) and for the OLJ-PB and OLJ-MB samples (bottom). Representative close-up images of the observed crack-events are pointed out with red arrows and squares. The cracks were counted manually through the analysis software of the optical microscope AXIO Imager from Zeiss.

*C. US Coil MQXFA AUP 213 from Magnet MQXFA08*

Coil AUP 213 from quadrupole magnet MQXFA08 experienced several detraining quenches below nominal current during cold testing. The voltage taps and quench antenna data pointed to the same location of the limiting quenches in the previously examined coil AUP 214. In addition, analysis of MQXFA08 assembly data suggested the same failure mechanism found in MQXFA07 [25].

Based on the outcome of the inspections performed on coil AUP 214, the inspection of the S-W transition was carried out by cutting the 4 longitudinal samples (2 at the LJ side, 2 at the OLJ side) containing the 4 cables facing these transitions in the coil connection side. As a first step, the OLJ side of the AUP 213 was examined in the same way as coil AUP 214 was examined: the PB and MB longitudinal samples were polished down to the midplane of the first row of strands facing the S-W transition and then observed by optical microscopy. Figure 9 shows that 105 transverse crack-events were identified over the PB longitudinal cross-section. Also, this field of cracks (about 2 mm wide) was found to be at the interface between the resin and the Cu wedge as seen for AUP 214.

For the LJ side, a different approach was used. Both the PB and MB samples were further cut at the interface between the resin and the Cu wedge since it was expected to find transverse cracks at this location. After cutting, they were embedded in epoxy resin (see Figure 10). Instead of analysing the longitudinal cross-section of the cut samples, a deep Cu etching process was carried out on their transverse cross-sections. Since inspection of longitudinal cross-sections provides information about transverse cracks, it is expected that each crack identified in the longitudinal plane corresponds to a broken filament when inspecting the transverse cross-section after the Cu etching process. This was confirmed through the investigation of the LJ samples.

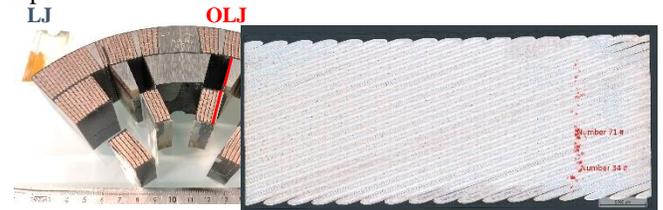

Fig. 9. MQXFA08 quadrupole magnet coil AUP 213: OLJ side inspection. The red marks on the prepared longitudinal cross-section represent all transverse crack-events (approx. 105) within a narrow field (2 mm wide) at the resin-Cu wedge interface.

Figure 11 presents the results obtained by etching the LJ-MB (top) and LJ-PB (bottom) transverse cross-sections.

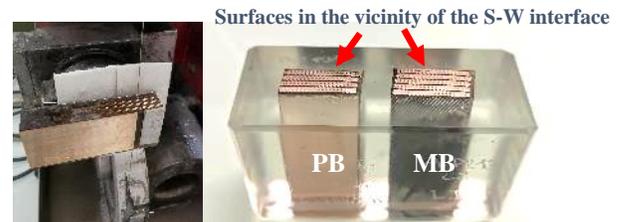

Fig. 10. MQXFA08 quadrupole magnet coil AUP 213: LJ-PB cut by DWS at the level of the resin-Cu wedge interface where the events are expected (left); LJ-PB and LJ-MB surfaces embedded in epoxy resin for the following Cu etching process (right).

In the LJ-MB cross-section 5 damaged strands and a total of 15 broken sub-elements were identified, whereas in the LJ-PB 15 damaged strands and a total of approx. 100 broken sub-elements were found. This outcome is very consistent with observations on the OLJ side, and on the previous specimens from AUP 214. Moreover, these results confirm that inspection of longitudinal cross-sections and observation of transverse cross-sections after deep Cu etching are methods that allow to highlight the same events from different cross-sections [9]. A given number of cracks observed in a longitudinal plane correspond to the same number of broken filaments visible after deep Cu etching in one or more damaged strands on the transverse cross-section.

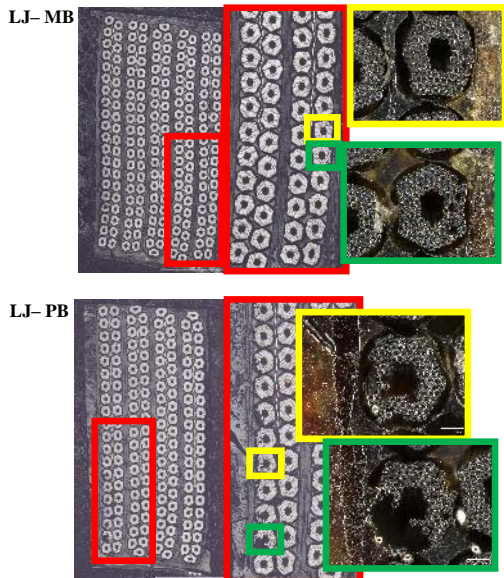

Fig. 11. MQXFA08 quadrupole magnet coil AUP 213: LJ-MB (top) and LJ-PB (bottom) after Cu etching. The red squares highlight a portion of the damaged area, and the images enclosed by yellow/green frames display a zoom in on the type of damage observed.

## IV. Conclusions

The application of the proposed examination method to performance-limiting coils allowed transversely broken damaged strands and broken sub-elements, likely associated to transport current degradation, to be univocally identified, at the positions of corresponding to quench start localisation. In particular, the deep Cu etching process performed on the CERN MQXFB quadrupole magnet coil CR108 revealed the presence of damaged strands featuring broken filaments at the top edge of the inner layer pole turn cable facing the titanium pole. This damage is not only observed in the coil central region where limiting quenches were found to originate, but also further from this zone towards the coil ends. The observed damaged strands were found to be always in the vicinity of the coil pole-to-pole transitions that represent discontinuity points. These transitions are thus key points to be explored in future inspections to help identifying at which stage of the coil fabrication or operation the degradation occurred.

The inspection of the US MQXFA quadrupole magnet coils AUP 214 and AUP 213, revealed in both cases a dense, narrow field of cracks at a specific longitudinal location between coil wedges and end spacers, at the interface between the resin and the Cu wedge. This location corresponds to a stress concentration point identified through finite element simulations, resulting from some non-conformities during coil and magnet assembly, partially attributed to Covid mitigation requirements [25]. The use of both Cu etching and longitudinal inspection for the AUP 213 investigation clearly proved that these techniques allow to reveal the same phenomenon from two different perpendicular planes: a transverse crack observed through the longitudinal cross-section corresponds to a broken filament identified after deep Cu etching.

The presented method has proven to be efficient in identifying and characterising the nature and extent of $Nb_3Sn$ sub-element breakage in relation to quench start localization in performance-limiting accelerator magnet coils. These findings enable to conclude that performance limitation on MQXF magnets can be traced back to extended and/or severe conductor damage at a location that corresponds almost systematically to transitions or singularities such as Titanium pole ends (broken filaments) for MQXFB coils or end spacers - wedge transitions (large field of cracks, broken filaments) for MQXFA coils. This result is consistent with previous observations on 11T coils [9] where singularities were also identified in the vicinity of spacers, under the form of bulged or popped in/out strand events. This also confirms that the performance degradation or limitation of HL-LHC $Nb_3Sn$ magnets is not an inherent feature of $Nb_3Sn$ technology but can be resolved by plain and good engineering practices, as demonstrated by the promising results obtained on the last MQXFBP3 prototype [26] and the most recent MQXFA magnets [25].

## V. Acknowledgements
The authors are grateful to the dedicated technical teams on both sides of the Atlantic who are contributing to the manufacture and test of the HL-LHC MQXF magnets. They are also grateful to the continuous support of the HL-LHC and AUP managements.

## References

[1] Apollinari, G. et al. "High-Luminosity Large Hadron Collider (HL-LHC). Technical Design Report V. 0.1." No. CERN-2017-007-M; FERMILAB-DESIGN-2017-03. Fermi National Accelerator Lab. (FNAL), Batavia, IL (United States), 2017.

[2] Rossi, L. and Brüning, O., 2012. "High luminosity large hadron collider: A description for the European strategy preparatory group" (No. CERN-ATS-2012-236).

[3] Ballarino, A. and Bottura, L. "Targets for R&D on Nb3Sn Conductor for High Energy Physics," in IEEE Transactions on Applied Superconductivity, vol. 25, no. 3, pp. 1-6, June 2015, Art no. 6000906, doi: 10.1109/TASC.2015.2390149.

[4] FCC collaboration. "FCC-hh: the Hadron collider: future circular collider conceptual design report volume 3." European Physical Journal: Special Topics 228.4 (2019): 755-1107.

[5] Devred, A., et al. "Challenges and status of ITER conductor production." Superconductor Science and Technology 27.4 (2014): 044001.

[6] Mitchell, N. and Devred, A. "The ITER magnet system: Configuration and construction status." Fusion Engineering and Design 123 (2017): 17-25.

[7] Ekin, J. W. (1984). Strain effects in superconducting compounds. In Advances in Cryogenic Engineering Materials (pp. 823-836). Springer, Boston, MA.

[8] Stoynev, S. et al. "Analysis of Nb 3 Sn accelerator magnet training." IEEE Trans. Appl. Supercond. 29.5 (2019): 1-6.

[9] Santillana I. et al. "Advanced Examination of $Nb_3Sn$ Coils and Conductors for the LHC Luminosity Upgrade: A Methodology Based on Computed Tomography and Materialographic Analyses." Paper to be submitted to SUST.

[10] Balachandran, S. et al. "Metallographic analysis of 11 T dipole coils for high luminosity-large hadron collider (HL-LHC)." Superconductor Science and Technology 34.2 (2021): 025001.

[11] Sgobba, S. et al. "11T Phase 1: Material properties, CT and materialography." CERN internal report available upon request.

[12] Sanabria, C. et al. "Metallographic autopsies of full-scale ITER prototype cable-in-conduit conductors after full cyclic testing in SULTAN: II. Significant reduction of strand movement and strand damage in short

twist pitch CICCs." Superconductor Science and Technology 28.12 (2015): 125003.

[13] Ebermann, P. et al. "Irreversible degradation of Nb3Sn Rutherford cables due to transverse compressive stress at room temperature." Superconductor Science and Technology 31.6 (2018): 065009.

[14] Ebermann, P. et al. "Relevance of the irreversible degradation of superconducting Nb3Sn wires and cables caused by transverse stress at room temperature within the FCC study at CERN." Diss. Wien, 2019.

[15] Puthran, K., et al. "Onset of mechanical degradation due to transverse compressive stress in Nb3Sn Rutherford cables", submitted to IEEE Trans. Appl. Supercond.

[16] Bagni, T. et al. "Formation and propagation of cracks in RRP Nb3Sn wires studied by deep learning applied to X-ray tomography." Superconductor Science and Technology (2022).

[17] Verhoeven, J. D., Gibson, E. D., and Laabs, F. C. "Filament contact within situ Nb3Sn superconducting wire." Journal of materials science 19.8 (1984): 2459-2464.

[18] Sue, J. J., et al. "Effect of Ta Additions upon in situ Prepared $Nb_3Sn$-Cu Superconducting Wire." Metallurgical Transactions A 15.2 (1984): 283-286.

[19] Todesco, E., et al. "The high luminosity LHC interaction region magnets towards series production." Superconductor Science and Technology 34.5 (2021): 053001.

[20] Izquierdo Bermudez, S. et al. "Progress in the development of the $Nb_3Sn$ MQXFB quadrupole for the HiLumi upgrade of the LHC." IEEE Trans. Appl. Supercond. 31.5 (2021): 1-7.

[21] Mangiarotti, F. J. et al. "Power test of the first two HL-LHC insertion quadrupole magnets built at CERN." IEEE Trans. Appl. Supercond. 32.6 (2022): 1-5.

[22] Ambrosio, G. et al., "Lessons Learned from the Prototypes of the MQXFA Low-Beta Quadrupoles for HL-LHC and Status of Production in the US" IEEE Trans. Appl. Supercond., 31 (2021) 4001105.

[23] Ambrosio, G. et al. "MQXFA final design report." arXiv preprint arXiv:2203.06723 (2022).

[24] Ambrosio, G. et al., "Analysis of MQXFA07 Test Non-Conformity." US-HiLumi-doc-4293 and CERN EDMS 2777612, available upon request.

[25] Ambrosio, G. et al. "Challenges and Lessons Learned from fabrication, test and analysis of 8 MQXFA Low Beta Quadrupole magnets for HL-LHC." Submitted to IEEE Trans. Appl. Supercond.

[26] Izquierdo Bermudez, S. et al. "Status of the MQXFB Nb3Sn quadrupoles for the HL-LHC." Submitted to IEEE Trans. Appl. Supercond.